# ICT Green Governance: new generation model based on Corporate Social Responsibility and Green IT

**Rachid Hba[1*], Abdellah El Manouar[2]**

1 ENSIAS Engineering School, Mohammed V University In Rabat

2 ENSIAS Engineering School, Mohammed V University In Rabat

*Corresponding author: Rachid Hba rachid_hba@um5.ac.ma

**Abstract**
The strategy of sustainable development in the governance of information and communication technology (ICT) is a sector of advanced research that leads to rising challenges posed by social and environmental requirements in the implementation and establishment of the governance strategy. This paper offers new generation governance model that we call "ICT Green Governance". The proposed framework provides an original model based on the Corporate Social Responsibility (CSR) concept and Green IT strategy. Facing increasing pressure from stakeholders, the model offers a new vision of ICT governance to ensure effective and efficient use of ICT in enabling an enterprise to achieve its goals. We present here the relevance of our model, on the basis of a literature review, and provide guidelines and principles for effective ICT governance in the way of sustainable development, in order to improve the economic, social and environmental performance of companies.

## INTRODUCTION

The ICT governancein the global ICT management of the companies is impacteddue to the technical and economic evolution. In consequence, on the socio-economic side, the results are substantial: oriented sustainable development models are completely open to the new forms of action, and potentially prepared to detect and put into action the good practices of ICT governance strategy. Several are the companies that recognize the potential of the Green IT (green information technologies), the Green IT make reference to ICT with low environmental impact through the reduction of carbon footprint, energy consumption and associated costs, throughout the life cycle of hardware, software and ICT related services, they also recognize the effect of Green IT on their organizations, on their overall ICT strategies, as well as, on the sustainable development or CSR strategies [8, 9]. The big challenge is to associate the ICT strategy with Green IT to provide new managerial and organizational perspectives, and green market opportunities.
Green IT and CSR activities are a recent concept which is seriously taken [3]; it is not perceived any more like fashion trend or a simple marketing label. In this context a new and great reflection is conducted on the CSR and Green IT management model, beyond marketing and environmental aspects [7]. In fact, the innovation activity associated with an ecological concern becomes a must to make the difference within a competitive and globalized environmental context. Thus, was born a new eco-responsible approach of ICT governance which is declined so much on the processes of the company, that on the comprehensive ICT strategy.

This approach leads us to speak about the green governance that we can qualify as new generation managerial model, which aims at the alignment of CSR strategy and Green IT with the global ICT governance strategy to reach the sustainable ICT governance [2]. Thus, we postulate that the green governance based on Green IT and the CSR is an alternative of conventional governance that will address a new perspective of structuring of organization and deep changes in the practices of ICT management.

To validate this assumption, our work will focus as well on two parties: we will first elucidate the concepts of green governance such as an agile model and horizontal managerial culture in the opposite of current models based on a hierarchically pyramidal relation and modes of subordinate collaboration, and we will outline the theoretical foundations of this research into sustainable management and eco-responsibility for the ICT governance. Secondly, we will expose the approach of the Corporate Sustainability and Responsibility (CSR) and the adoption of the Green IT practices like new model of good governance practices that proposes a 360 degree approach in terms of decision making criteria and offers a transverse dimension for ICT governance. This approach will invite us to develop the green governance as a new generation model of management thanks to the adoption of Green IT and CSR practices, this model has two dimensions: the Green IT and CSR, and governance, and how a strategy Green IT can give place to a strategy corporate and constitute the core of a total CSR strategy.

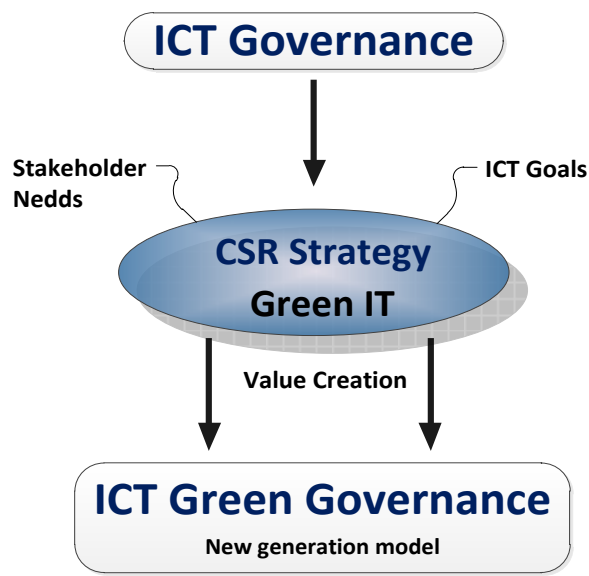

Figure 1. Synthesis of the problems

**I LITERATURE REVIEW**
**1.1 Dominant approaches of ICT governance strategy**
Face to dynamism to increase corporate policy and ICT governance strategy, several studies from the sixties have focused on aspects impactingglobal performancewithin companies with direct effect on strategic choice and competitive advantage. Further new processing logic, a deep technological and societal change has allowed researchers and engineers to rethink the conventional governance model by seeking innovation and the value creation in the organizations and business models in order to developefficiency and competitiveness through the ICT revolution that offers an innovation manner and growth opportunities.

At the moment, ICT offers a continuous flow of knowledge creation and creative managerial practices of sensethrough their adaptability and their flexibility, and upgrade the flow of information between various stakeholders (recipients) [15], in a many to many way, on the social and economic as well as environmental aspect. Owing to, their technological prowess and their extended and cross scope, ICT are able to bring a plus-value in the establishment of the managerial and strategic policies of the companies by taking into account the institutional requirements and taking care to conserve legitimacy via a socially constant control.

Information and communication technology governance or "ICT Governance" is a sub field of corporate governance focused on information and communication technology [33]. The aim of this discipline is interested primarily in the risk management, resources and investments optimization, in total alignment with company goals, to improve ICT performance and create value [35] via specifying the decision rights and accountability framework.

The field of ICT governance defines a management analysis framework through appropriate organizational structures that allow controlling and ensuring that ICTs contribute to the management strategy of the company [47] and value creation through piloting devices for compliance and control of IT processes.As we have emphasized, ICT governance and a search field that draws the principle of global governance strategy within the organization, thus it is directly influenced by other areas of corporate governance that are financial governance and corporate social and environmental governance[49].The goal of our research is henceforth, to explore and develop governance based on eco-responsible practices of ICT in a global perspective taking into account the challenges of sustainable development.View the wave of growing integration of the CSR concerns in organizations, the need to move towards a new way of practicing governance by setting a target of reducing the environmental, financial and social footprint of ICT throughout the life cycle of governance processes [49] and in a contribution approach to the overall sustainable development strategy of the company.

According to the COBIT (Control Objectives for Information and related Technology)developed from the ITSEC monitoring technical reference that was first published in May 1990 (Information Technology Security Evaluation Criteria), ICT governance is a kind of structure ofrelationships and processes to pilot and control the business to achievedesired goals by generating value while finding the right balance between the benefits of ICT management and processes establishment and risk control.The same definition was reported by COSO published in the USA in September 1992, COSO is the abbreviated acronym for Committee of Sponsoring Organizations of the Treadway Commission (the model objective isbased on the Internal Control Integrated Framework) [48].

The governance model designed by COBIT is based on a process-oriented approach that was developed in 1994 and published in 1996 by the ISACA association (Information Systems Audit and Control Association). Its aim is to help decision makers to manage risk, reliability ICT compliance, and investment via a control framework [47] which includes four fields; planning and organization, acquisition and installation, delivery and support and monitoring.

The COBIT 5 version speaks a little of the concept of "Policies and Culture", but there is no explicit indication of Green IT practices and social responsibility in the policy development process or computer activities, we can talk about what is a missed opportunity. However the concept of Policies and Culture is the area where Green IT was born and evolved in organizations.

Another contribution was led by AFNOR (French Association for Standardization) [50, 51] by the establishment of BP Z67-320 repository of best practices; this repository provides a governance approach and practical recommendations on the eco-responsible information systems. Especially, reducing the energy consumption of ICT and minimizing their environmental impact. This contribution provides organizations best practices for creating green value and support innovation in the field of sustainable development.

The governance approach presented will integrate the concept of sustainability as a key decision-making criterion, and ensure alignment of the Green IT initiatives with the overall strategy of the company and also to identify areas for improvement to ensure continued innovation of ICT, using indicators for sustainable development [50, 51].

**1.2 Corporate Sustainability and Responsibility (CSR)**
Corporate social responsibility (CSR) is a theme that was discussed by Henry Ford in 1920 and he wrote "If you try to run a business solely on profit, then it will also die because it will not have more reason to be". After several years the concept was defined in 1953 by Bowen and was enhanced following the Brundtland Report in 1987 and the Rio Summit in 1992, with an aim of looking further into the reflection on the regulation of the globalization phenomenon and multiplication of policies and economic interactions [13]. The "Global Compact" pact was concluded at the World Economic Forum in Davos on January 31st, 1999 following the suggestion from the Secretary-General of the United Nations Kofi Annan; this pact stipulates the stakes of the CSR based on sustainable development and common values which will enhance the investment of the human capital [17].

The CSR became an engine of transformation towards a new form of governance of the company (article 49, "Grenelle de l'environnement I" law). The United Nations, the Economic Co-operation and Development (OECD), the European commission and ISO consider "the CSR is the contribution of the companies to the stakes of sustainable development, and their responsibility towards the social and environmental impacts of their activities", this definition joined the representation into 1997 of John Elkington by Triple Bottom Line (TBL) which shows CSR stakes of sustainable development [22]:

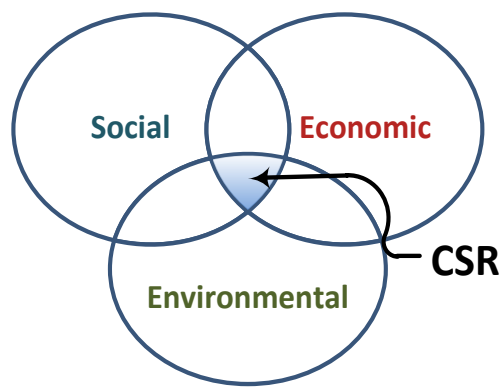

Figure 2. Venn diagram of the Corporate Sustainability and Responsibility

The concept of the CSR is a strategic lever of action which fits in the sustainable and which includes all the elementsbeing able to create a plus-value of the business on the long term [8]. The progressive CSR approach is a sustained approach and controlled by a societal partnership contract which frames the interactions between the actors giving a new design to the organization [9] and the governance of the company.

**1.3 Green IT**
The biggest challenge facing the environment today is global warming, caused by carbon emissions. According to the Energy Information Administrationreport, About 98% of $CO_2$emissions can be directly attributed to the energy consumption. Today, any companies are speaking openly about a need to operate in a greenmode, publishing sustainability principles for environmental practices on their corporate identity. In addition, many companies are now paying some kind of carbon duty for the resources they use and the environmental impact of services and the products they manufacture, thus, reducing the energy consumed can have a real financial recovery.

After the oil crisis baptized "First Oil Shock" in 1973 and the "Second Oil Shock" in 1979, the government of the United States launched in 1992 the "Energy Star" program for achieving energy saving in the ICT and data-processing field, this phase is also marked by the publication in 1993 of a first eco-label in the Green IT domain. The "Energy Star" program was initiated by the Environmental Protection Agency (EPA) after the Rio de Janeiro Earth Summit which joined together nearly 110 Heads of States and governments and represented by 178 countries. This summit was marked by the adoption of the "Rio Declaration on Environment and Development" consisted of 27 principles, thus concretizing the concept of sustainable development, an essential step in the birth of the Green IT concept.

The concept of Green IT was a great evolution from of an objective of reducing ICT impact on the company organization (organization footprint) and on the environment (carbon accounting) [16] to consideration of the impacts on the society and others stakeholders. Green IT was increasingly become a lever of CSR strategy providing strategic management and CSR reporting techniques [14], and for the causal relationship, the Green IT is the countenance of CSR [1]. The "Smart 2020" [34] study conducted by professionals of sustainable development has a focus on Green IT leverage on the company's processes thanks to its positive contribution to sustainable development. One of the key numbers presented by this study shows that Green IT will allow, by 2020, to make economies in terms of $CO_2$ emission of about 7800 million tons.

With the publication of the ISO26000 standard and the "Grenelle de l'environnement II" law, the Green IT became more ambitious and thus offers a tool for strategic and operational management of CSR stakes, allowing reduction of the ecological footprint of ICT [3, 16]. The evolution of green IT approach and its related fields that contribute to the company's involvement in CSR strategy, showed two levels or phases of Green IT maturity: we distinguish the Green IT 1.0 (Green For IT) and the Green IT 2.0 (IT For Green):

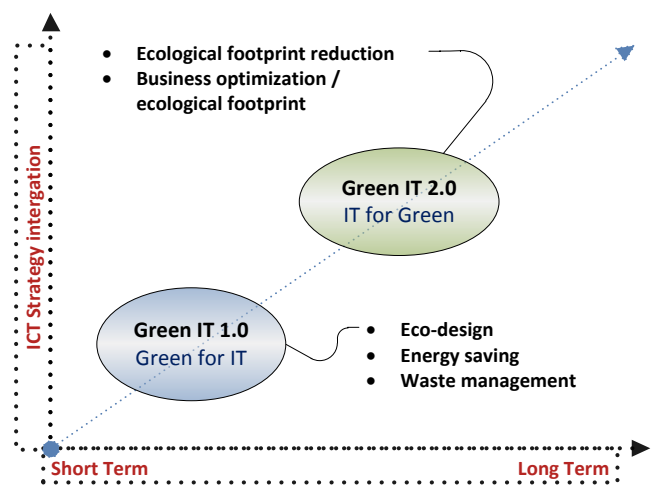

Figure 3. Evolution of Green IT

The purposes shared by these two types are generally, the energy effectiveness and saving, reduction of the company footprint and economic competitiveness. The Green IT 1.0 proposes good practices of use of computer material in order to minimize the environmental impact of ICT (eco-responsible ICT). On the other hand, the Green IT 2.0 takes an interest, through the ICT, in reduction of ecological footprint of the company (CSR stakes) it also allows the optimization of business processes and organization of the supply chain (Green Supply Chain).

**1.4 Stakeholder theory**
During last years, a very specific attention is drawn by the neo-institutionalists on the concepts of stakeholder's paradigm [34]. The researches carried out in the academic field have studied the company as being an organization or an institution, having potential and values towards all stakeholders [38, 39]. The stake goes beyond concerns of shareholders benefits to the broader surrounding aspects.

Few, are the studies which were made on this concept of stakeholders, the literature misses works exposing the related subject and theories. However, the stakeholder's theory is exposed according to a model of organizational relations between the company and its economic and environmental partners [40]. Its objective is to propose a tool for the improvement of the means of subsistence and durability [41]. Thus, it is regarded in certain cases as an obvious result of the social sciences evolution, in other cases; it is seen as a simple framework directed by the principles of morality [42].

With the emergence of Green IT and CSR, the research on the topic of stakeholders [43] provides a reference framework to theoretically develop the CSR which will incite us to explore the common axes that could lead us to a new theory or a new concept for use on many levels and for purposes of broadening the scope of its application.

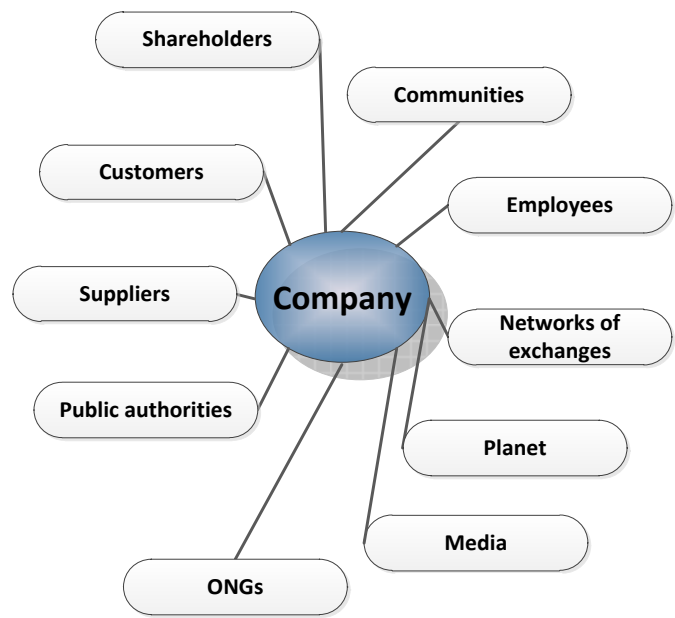

Figure 4. Stakeholder theory diagram

**II ICT GREEN GOVERNANCE: NEW GENERATION MODEL**
ICTgovernance was introduced in companies in a context where from a point of view; the automation of activities and processes has become a key element within the company and from another point of view, where decision makers do not see how ICT can bring plus-value, effectiveness and performance in the organization. We can as well talk about ICT governance and therefore theoretical models and organizational and managerial practices. It is also in the interest of the transformation and innovation of techniques developed for the control and monitoring of ICT, to seek and explore new models for effective corporate governance, which requires strong management skills to make important decisions and provide leadership.The main objective is to propose a model for mastering ICT governance over time. This framework should be based on a set of "best practices" that are incorporated within the governance structure within the organization [35, 36], and may take the form of changing the architecture of business processes [46].

As we explained in the introduction, with the appearance of the concept of sustainable development, the concerns and the engagement of the organizations in the business of ICT management are at the same time complex and improved and allowed to evolve missions and managerial competences of the decision makers [10, 11]. Thus, environmental dimension and responsibility values endeavoured the companies to choose innovating practices in order to satisfy the sustainable development requirements.

ICT governance includesthe principles and the procedures as well as the organizational management and structures which ensure that a firm's IT sustains and extends objectives and the organization's strategies, that the IT's resources are used in a responsible way, and that its risks are attenuated and controlled [37, 38]. To reach the suggested outcomes, it is recommended to apply various mechanisms which, in this article, will be referred to as ICT green governance model. The following Frameworkhave been described in the literature: ICT strategy, ICT conformity and risk management, ICT compliance management, ICT performance management, ICT service management, IT sourcing management, ICT portfolio management.

Our model allows the sustainability of the resources (CSR impact on the human resources management and processes) [30, 39]. It is, therefore, an effective and efficient organizational framework of the balance of the sustainable resources.

Given that ICT incorporates several actors in the production and the information processing and face to the difficulties to define a conceptual framework of these increasingly complex systems, the partnership (management of the stakeholders impact on the organization) and contractual approach (use of ICT for goal of organizational balances) of this model contribute to the enhancement of organizational and managerial architectures of the ICT business, as well on the creation of information as on its processing [11].

In our research we use the green governance as a model or management Framework to align conventional management to sustainable development concerns and the ICT Governance. Our major contribution is the use of a new strategy that combines governance and alignment as part of an integrated approach of Green IT and CSR. The Framework consists of two pillars or dimensions of good practices:

- Green IT and CSR Approach
- ICT Governance

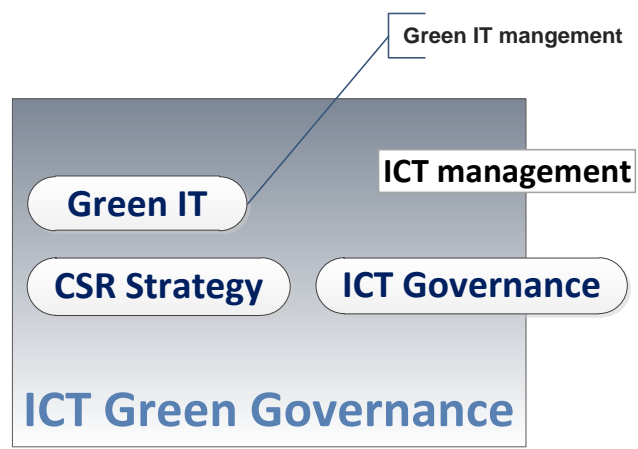

Figure 5. ICT Green Governance model

Our ICT green governance model offers to the managers two visions of decision making [13] to encourage desirable behavior in the use of ICT; an internal vision which consists in improving the efficiency and in minimizing the costs of the processes, and an external vision, when ICT is used in order to create a unique customer value [10] and this vision allows to reach goals of sustainable development towards various stakeholders thus to improve business competitiveness and the leadership of the company:

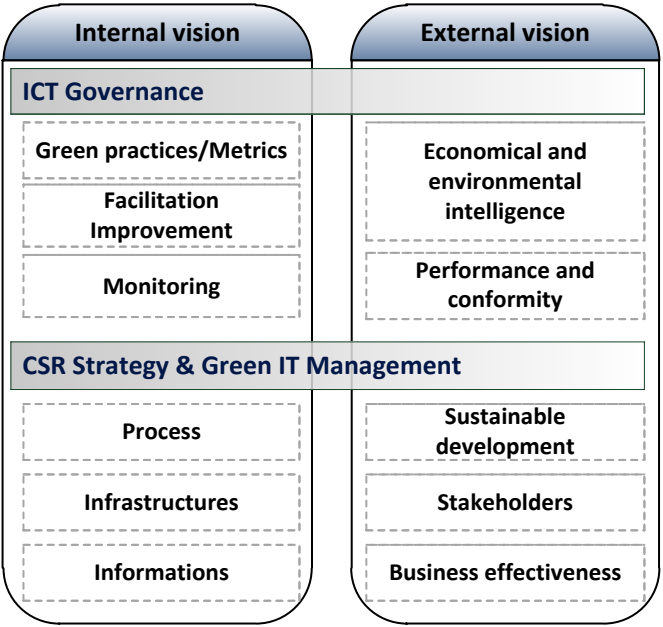

Fig. 5.Internal and external vision of the ICT green governance

The ICTgreen governance model aims to the identification, with low costs [21], of sources of competitive advantage and thus aligns the internal vision of management with the external strategy which impacts other stakeholders. To achieve these objectives, our model offers an eco-responsible approach focused on two dimensions: Green IT and CSR, and ICT strategic governance. This helps to develop management strategies that can improve low-impact sustainability processes and business operations. The green governance provides a comprehensive framework that will help companies create optimal value thanks to the two mentioned dimensions, thus maintaining the balance between the goals of various internal and external [20] stakeholders and optimization levels of risk and the use of ICT.

To deliver value to stakeholders, the company needs good Green governance that evaluates and control, in a holistic manner, activities related to the use of ICT. This axis of our model is represented by the deployment of CSR type of governance authorities which provide indicators and dashboards management activities related to sustainable development for effective and sustainable management of ICT. The goal is to meet the requirements of lawful and contractual performance and conformity with respect to the other stakeholders and to make sure "that all the directives and internal and external regulations of conformity, are considered and treated in a suitable way".

The Green IT and CSR approach in the context of the governance which is exposed on the model is conceptualized by a approach eco-responsible which obliges the managers to integrate durability in the instruments of governance, recommendation which must be applied in the various mechanisms of governance; ICT strategy, control risks and conformity, performance, resources management and economic and environmental intelligence, also "the infrastructure must be managed in the most effective manner, at the same time from the environmental and financial point of view" [4]. Thus, the integration of the notions of Green IT and CSR in the ICT governance implies necessarily a sustainable governance, "a sustainable IT strategy should be aligned on the sustainability strategy on the scale of the company" [5] This sustainable governance allows identifying and structuring the requests of the stakeholders by taking heed of the economic, environmental and societal concerns. Consequently, the methods and the criteria of conventional governance are reinforced by the CSR and Green IT factors.

**Conclusion**

Our work in this paper exposes the state of the art of the green governance, which we have qualified as new generation managerial model. The conceptual model, presented here, is a part of a research work which is in progress. It establishes a basis to pilot a reflection on the new generation ICT management systems based on CSR approach and Green IT strategy, and aims to enhance research in the field of responsible ICT management. Our major contribution is the use of a new strategy that incorporates governance as part of an integrated approach of Green IT and CSR strategy, which aims to ICT management differentiation based on sustainability-related value.

It is therefore too early to draw conclusions; however, the objective of our model is to contribute to the perception of an ICT responsible management, in order to provide to the companies a roadmap to create new models of green business, thought on sustainable resources.